\documentclass{PoS}
\usepackage{wrapfig}
\PoS{PoS(HEP2005)339}

\title{Search for RPV Gaugino and Gravitino Production at HERA}

\ShortTitle{Search for RPV Gaugino and Gravitino Production at HERA}

\author{
 \speaker{Volker Adler}\\
 DESY Hamburg, Germany\\
 E-mail: \email{volker.adler@desy.de}
}
\author{
 On behalf of the H1 and ZEUS collaboration.\\
 Homepages:\\
 \href{http://www-h1.desy.de}{\tt www-h1.desy.de}\\
 \href{http://www-zeus.desy.de}{\tt www-zeus.desy.de}\\
}

\abstract{
 In R-parity violating supersymmetry a gaugino can be produced at HERA by the $t$-channel exchange of a selectron between the beam positron or electron and an initial quark from the beam proton.
 In the generic MSSM model the gaugino can decay in a cascade into two quarks and an (anti)lepton.
 For this model a search in the $e^\pm$-channel has been performed with the ZEUS detector using combined $e^\pm p$-data of $\mathcal{L}_{int}=121\,$pb$^{-1}$.
 No deviations from the Standard Model have been observed and exclusion limits on parameters of the MSSM have been set.
 Assuming the GMSB model with the gravitino as lightest supersymmetric particle and the neutralino as next-to-lightest supersymmetric particle, the neutralino decays into a gravitino and a high-energy photon.
 For this model searches have been performed with the H1 detector using $e^+p$-data of $\mathcal{L}_{int}=64.3\,$pb$^{-1}$ and $e^-p$-data of $\mathcal{L}_{int}=13.5\,$pb$^{-1}$ and with the ZEUS detector using $e^+p$-data of $\mathcal{L}_{int}=65.1\,$pb$^{-1}$.
 No deviations from the Standard Model have been observed and limits on the selectron and neutralino masses have been set.
 Additionally, limits on the R-parity violating Yukawa coupling have been derived in the search performed with the H1 detector.
}

\FullConference{
 International Europhysics Conference on High Energy Physics\\
 July 21st - 27th 2005\\
 Lisboa, Portugal
}

\begin{document}

 \section{Introduction}

 Supersymmetry (SUSY) \cite{Nilles:1983ge} is a promising extension to the Standard Model (SM) to answer to its open questions.
 This fermion-boson symmetry predicts a supersymmetric partner (sparticle) to each SM particle with identical mass and quantum numbers, but differing by $1/2$ in its spin.
 The minimal supersymmetric Standard Model (MSSM) introduces the minimum number of new parameters to incorporate SUSY into the Standard Model.
 SUSY has to be broken at present experimental energies, since no sparticles have been observed yet.
 Various SUSY breaking mechanisms are under discussion, and the choice is essential for the sparticle mass spectrum.
 In Gauge Mediated SUSY Breaking (GMSB) models, the gravitino is the lightest sparticle.

 SUSY introduces a new discrete, multiplicative quantum number, the R-parity
 \begin{displaymath}
  R_P=(-1)^{3B+L+2S}\,\mbox{,}
 \end{displaymath}
 which is $+1$ for SM particles and $-1$ for sparticles.
 The most generic Lagrangian allows R-parity violation (RPV), where at leading order the term
 \begin{displaymath}
  \mathcal{L}_{RPV} = \lambda_{ijk}L_iL_j\bar{E}_k+\lambda'_{ijk}L_iQ_j\bar{D}_k+\lambda''_{ijk}\bar{U}_i\bar{D}_j\bar{D}_k
 \end{displaymath}
 with the Yukawa couplings  $\lambda$, $\lambda'$, $\lambda''$ and the generation indices $i$, $j$, $k$ appears.
 In this framework, sparticles can be produced singly and decay to SM particles.

\begin{figure}[htbp]
  \begin{center}
   \includegraphics[width=50mm]{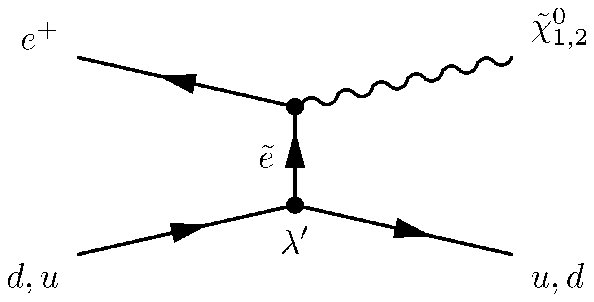}
   \hspace{10mm}
   \includegraphics[width=50mm]{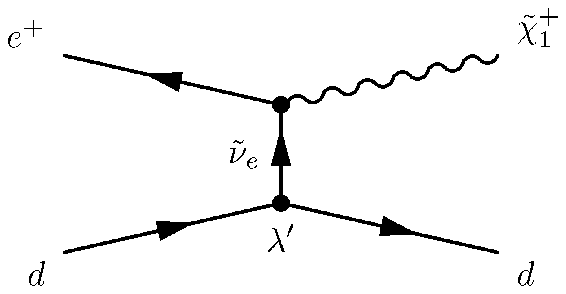}
   \\(a)\hspace{60mm}(b)
  \end{center}
  \caption{
   Gaugino production in $ep$-scattering at HERA.\protect\\
   The incoming lepton (here: $e^+$) interacts with a quark from the incoming proton via slepton exchange with production of a gaugino.
   The RPV Yukawa coupling $\lambda'$ takes place at the quark-slepton vertex.
   An exchanged selectron (a) can couple to both, $d$- and $u$-quark and results in an outgoing neutralino.
   An exchanged sneutrino (b) can couple only to $d$-quarks and results in an outgoing chargino.
  }
  \label{fmf_gaugino}
 \end{figure}

 As an $ep$-collider, HERA is highly applicable to probe for $\lambda'$.
 The possible RPV SUSY production mechanisms are $s$-channel squark production \cite{Fourletova:2005} and $t$-channel slepton exchange with gaugino production as shown in Fig. \ref{fmf_gaugino}.
 In the latter case, the cross section depends only on the masses of the gaugino and the exchanges slepton and not on the squark masses.

 \section{Search for RPV Gaugino Production}

\begin{wrapfigure}{l}{50mm}
  \includegraphics[width=50mm]{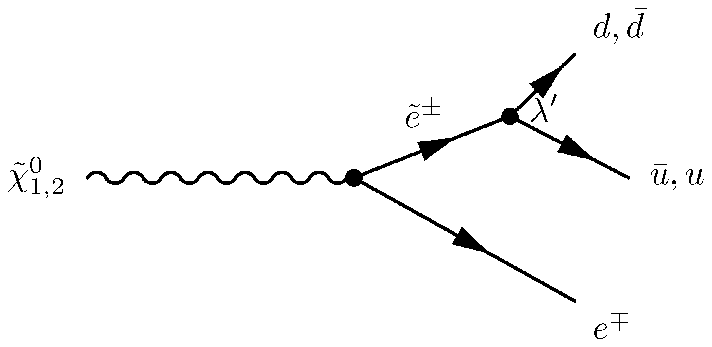}
  \caption{
   RPV gaugino decay.\protect\\
   This example shows the decay of a neutralino in the $e^\pm$-channel.
   The slepton from the gaugino decays further into (anti)quarks via an RPV Yukawa coupling $\lambda'$.
  }
  \label{fmf_decay}
 \end{wrapfigure}

 The ZEUS collaboration has searched for RPV gaugino production within the MSSM framework\footnote{No assumptions on the SUSY breaking mechanism are made.}.
 It is assumed that the RPV coupling is dominated by $\lambda_{111}'$.
 The gaugino decays into two (anti)quarks and a lepton from the first generation as shown in the example in Fig. \ref{fmf_decay}.
 The signal has been simulated with $\tan\beta=30$, large squark masses ($M_{\tilde{q}}=1\,$TeV) and degenerate slepton masses
 ($M_{\tilde{l}}=100\,$GeV).
 The coupling is assumed to be $\lambda_{111}'=1$.
 A parameter scan has been performed for the SUSY parameters $|\mu|<500\,$GeV and $M_2<250\,$GeV.

\begin{figure}[htbp]
  \begin{center}
   \includegraphics[width=65mm]{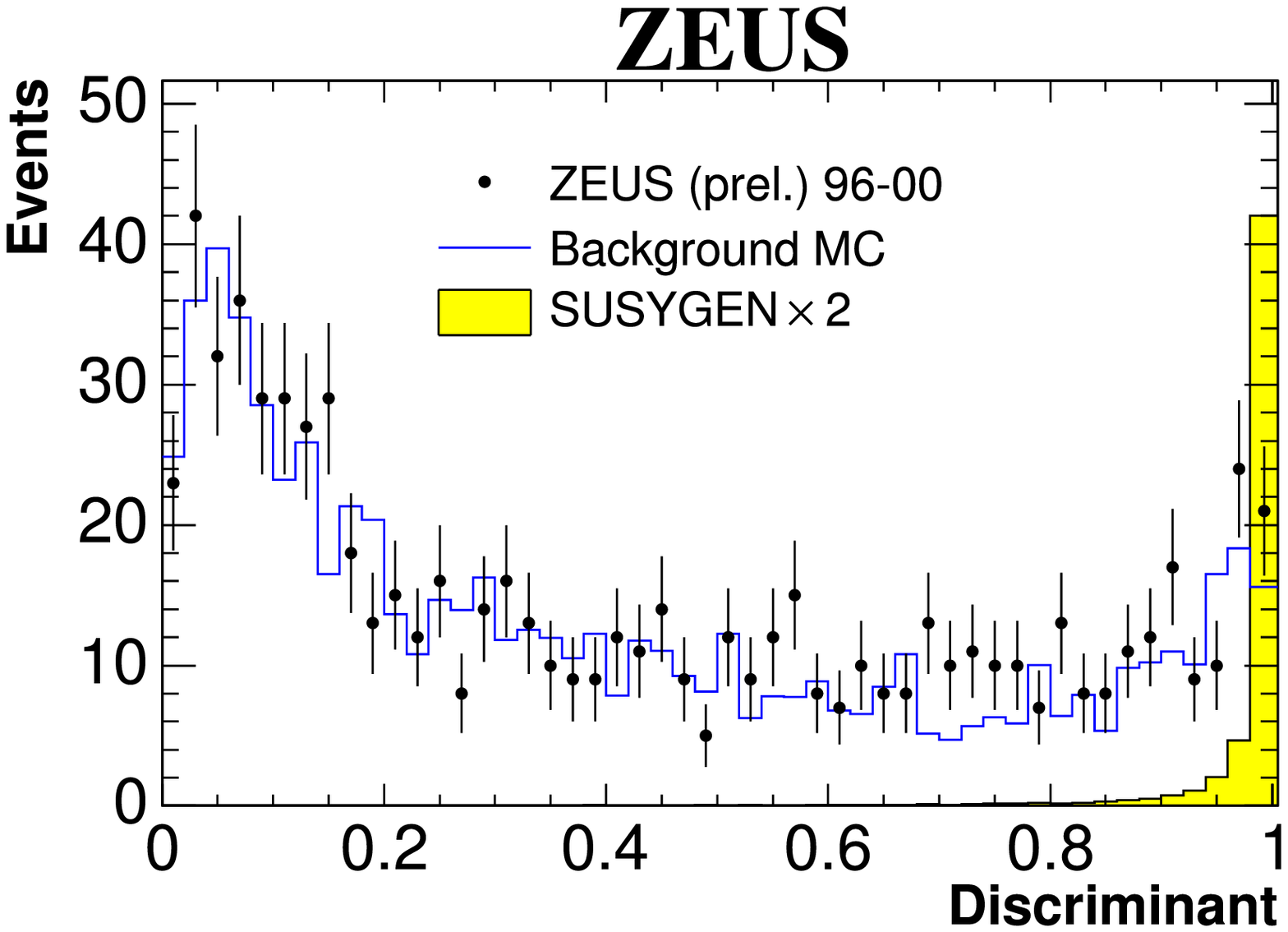}
   \includegraphics[width=65mm]{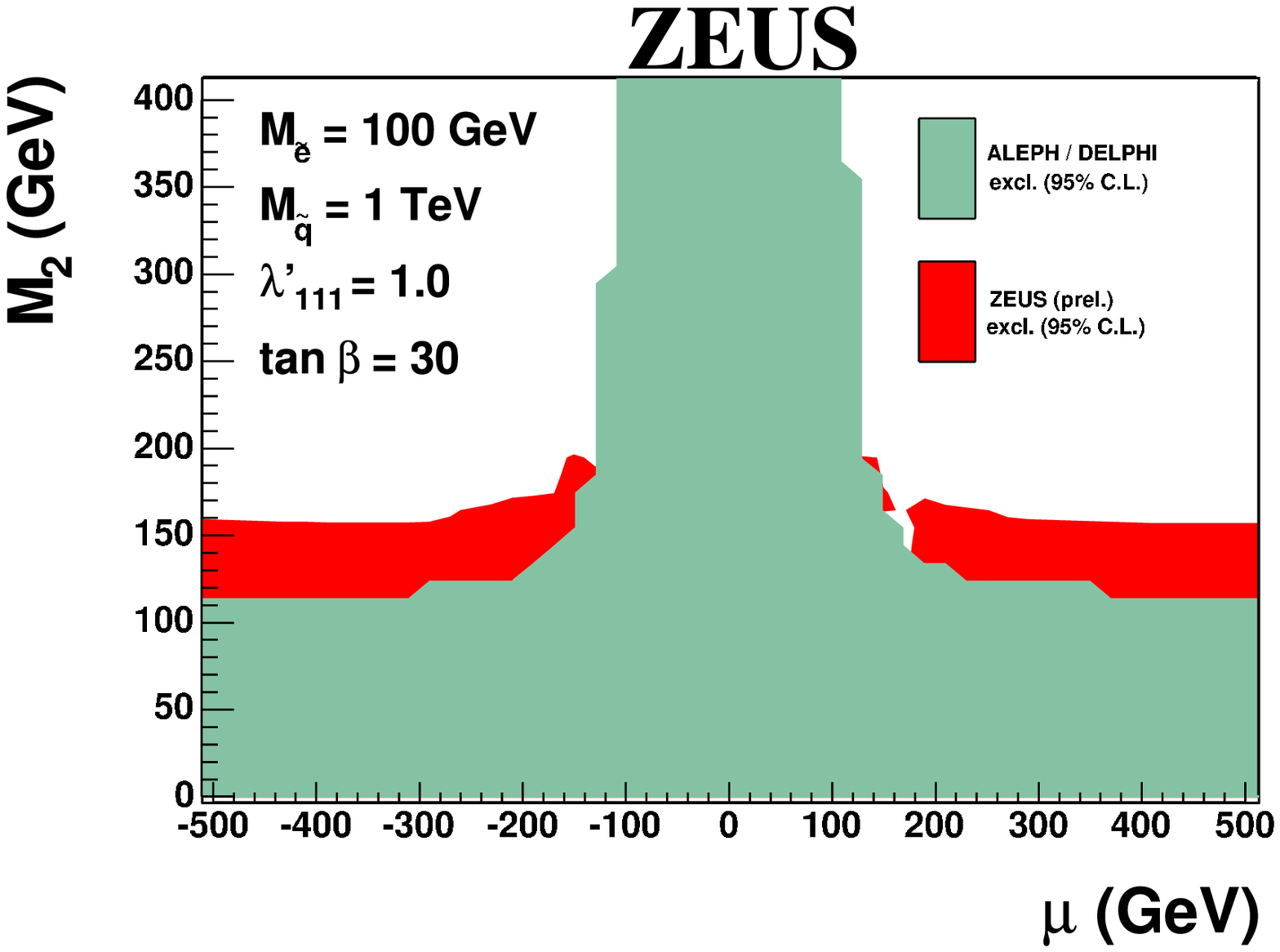}
   \\(a)\hspace{62mm}(b)
  \end{center}
  \caption{
   Discriminant and exclusion region at $95\%$ CL in the $M_2$-$\mu$ plane.\protect\\
   The distribution of the discriminant (a) is shown for the example of $\mu=-400\,$GeV and $M_2=150\,$GeV.
   No cut on the discriminant is applied.
   The signal efficiency multiplied by the branching ratio is $\approx 30\%$.
   The derived exclusion limits in the $M_2$-$\mu$ plane (b, red/dark area) are combined with former LEP results (green/light area).
  }
  \label{plot_ZEUS_gaugino}
 \end{figure}

 In the $e^\pm$-channel one expects events with high transverse energy, two jets with high transverse momenta and one $e^\pm$-candidate.
 The main SM background are neutral current deep inelastic scattering di-jet events.
 The found data events have been classified using a multi-variate discriminant method \cite{Carli:2002jp} with seven variables.
 As shown in Fig. \ref{plot_ZEUS_gaugino}, no deviations from the Standard Model have been observed in the high-discriminant region and limits on the scanned parameter space are derived.

 \section{Search for RPV Gravitino Production}

 The H1 and ZEUS collaboration have searched for RPV gravitino production within the GMSB framework with the neutralino as next-to-lightest sparticle, which decays under R-parity conservation via $\tilde{\chi}^0\to\gamma\tilde{G}$.
 The slepton and gaugino masses are not related by any model assumption, hence the mass difference
 \begin{displaymath}
  \Delta m = m(\tilde{e}) - m(\tilde{\chi}_1^0)
 \end{displaymath}
 can be small.
 The accessible couplings are $\lambda_{1j1}',j=1,2$ in $e^+p$-scattering\footnote{$j=3$ is suppressed due to the high top quark mass.} and $\lambda_{11k}',k=1,2,3$ in $e^-p$-scattering.
 One expects events with missing transverse momentum, one isolated high-energy photon and one jet.
 The main SM background are charged current deep inelastic scattering events with radiative or fake photons.

\begin{figure}[htbp]
  \begin{center}
   \includegraphics[width=58.1mm]{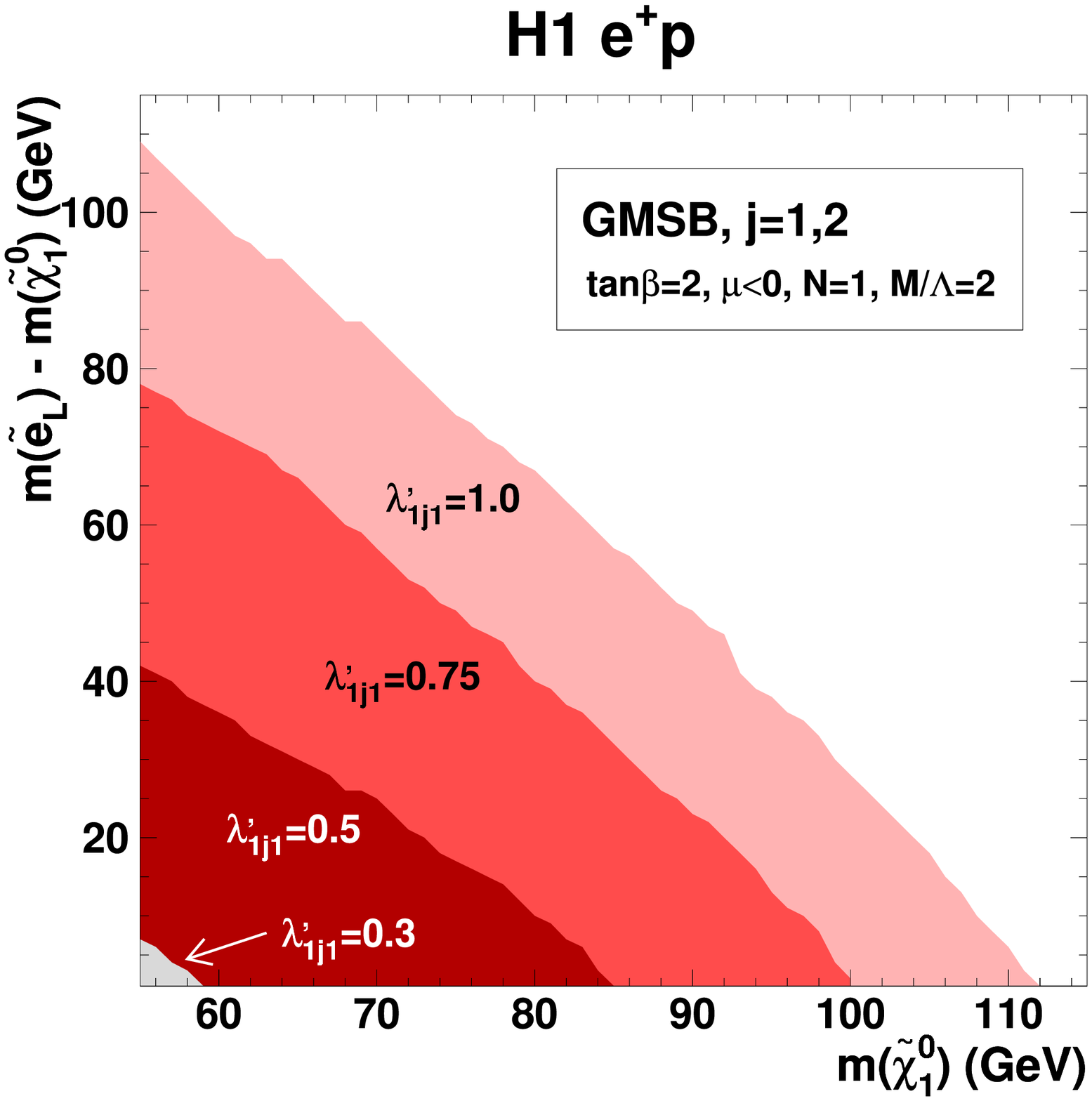}
   \hspace{3mm}
   \includegraphics[width=60mm]{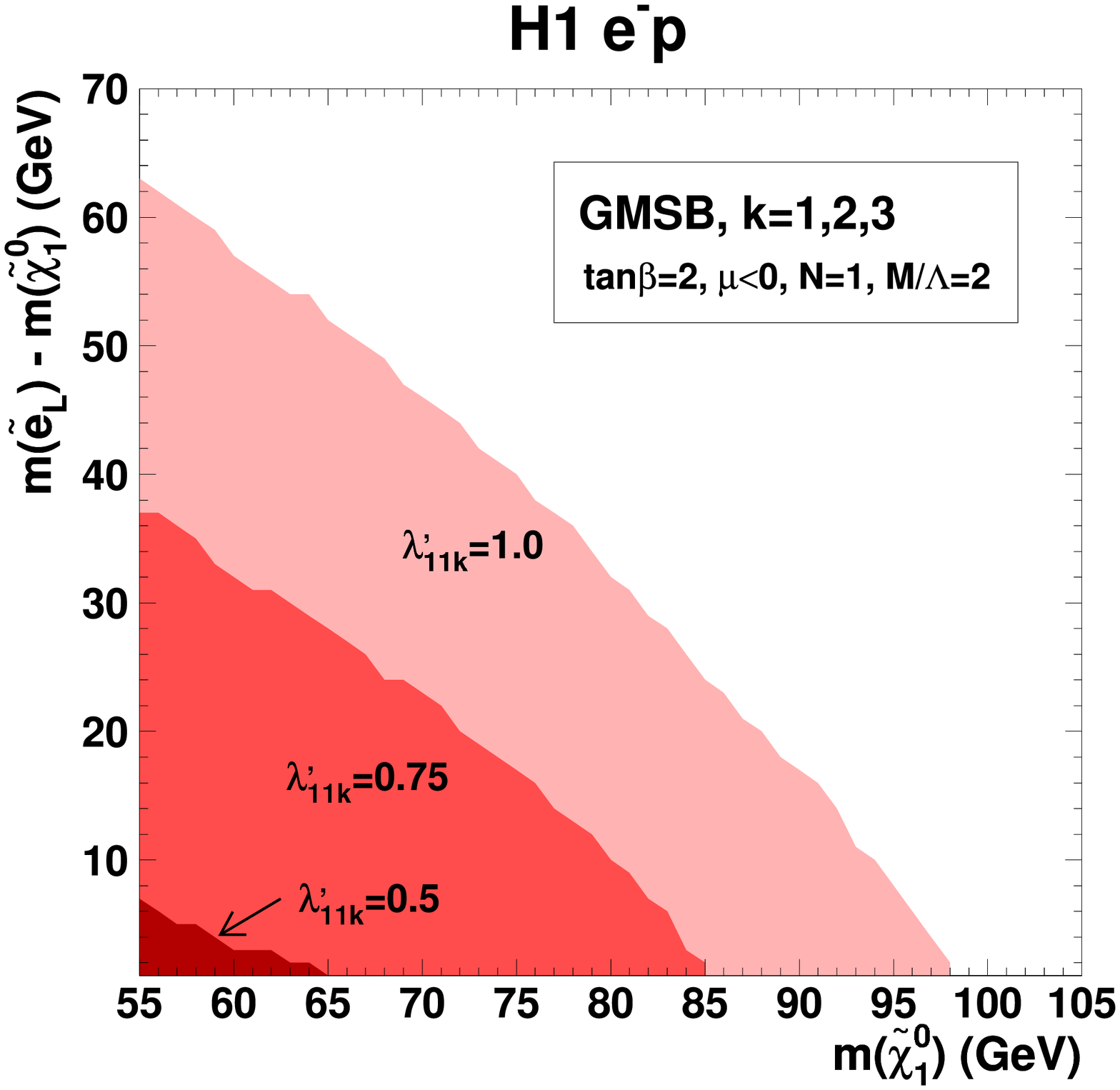}
   \\(a)\hspace{60.6mm}(b)\\
   \includegraphics[width=58.1mm]{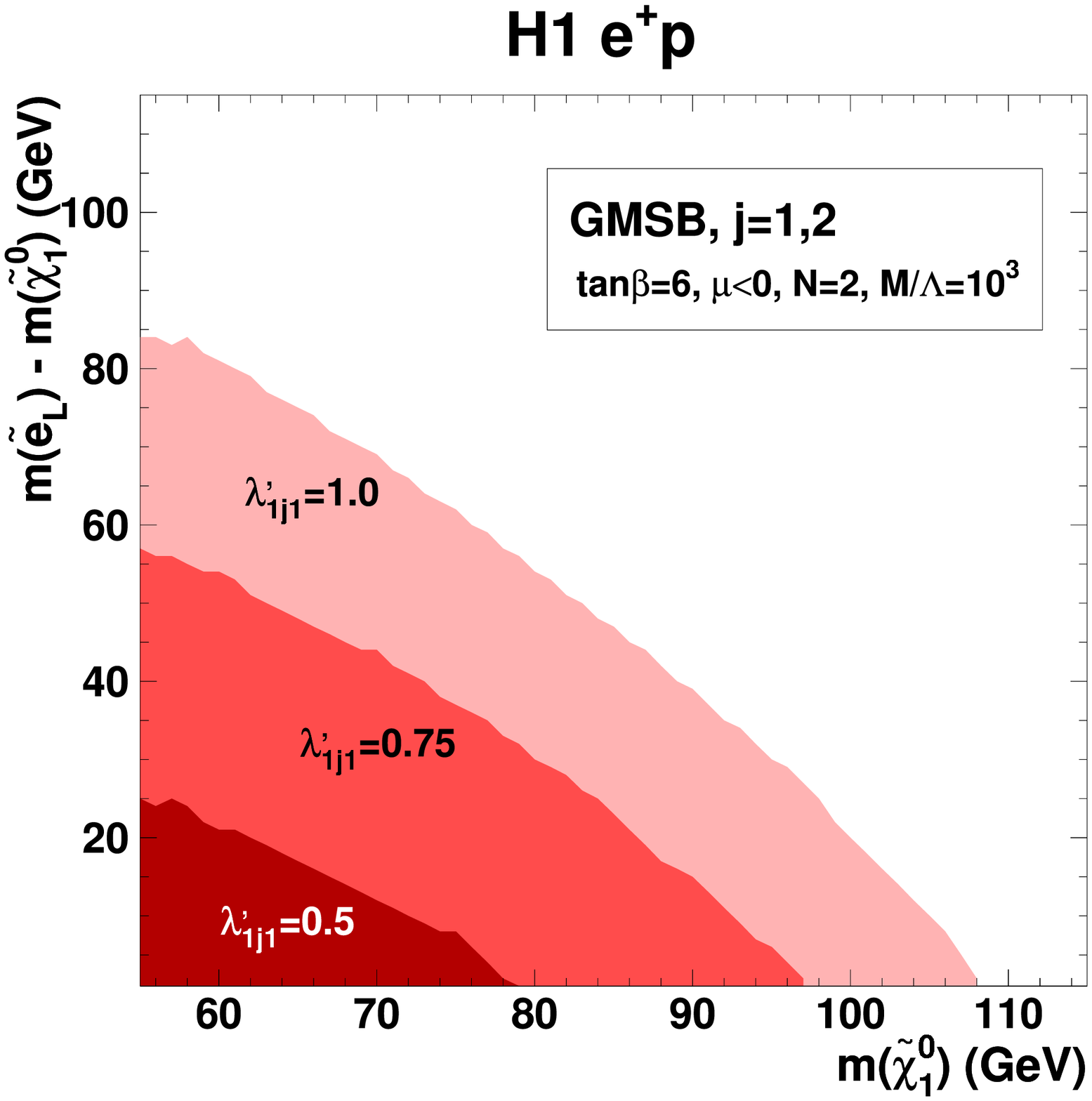}
   \hspace{3mm}
   \includegraphics[width=60mm]{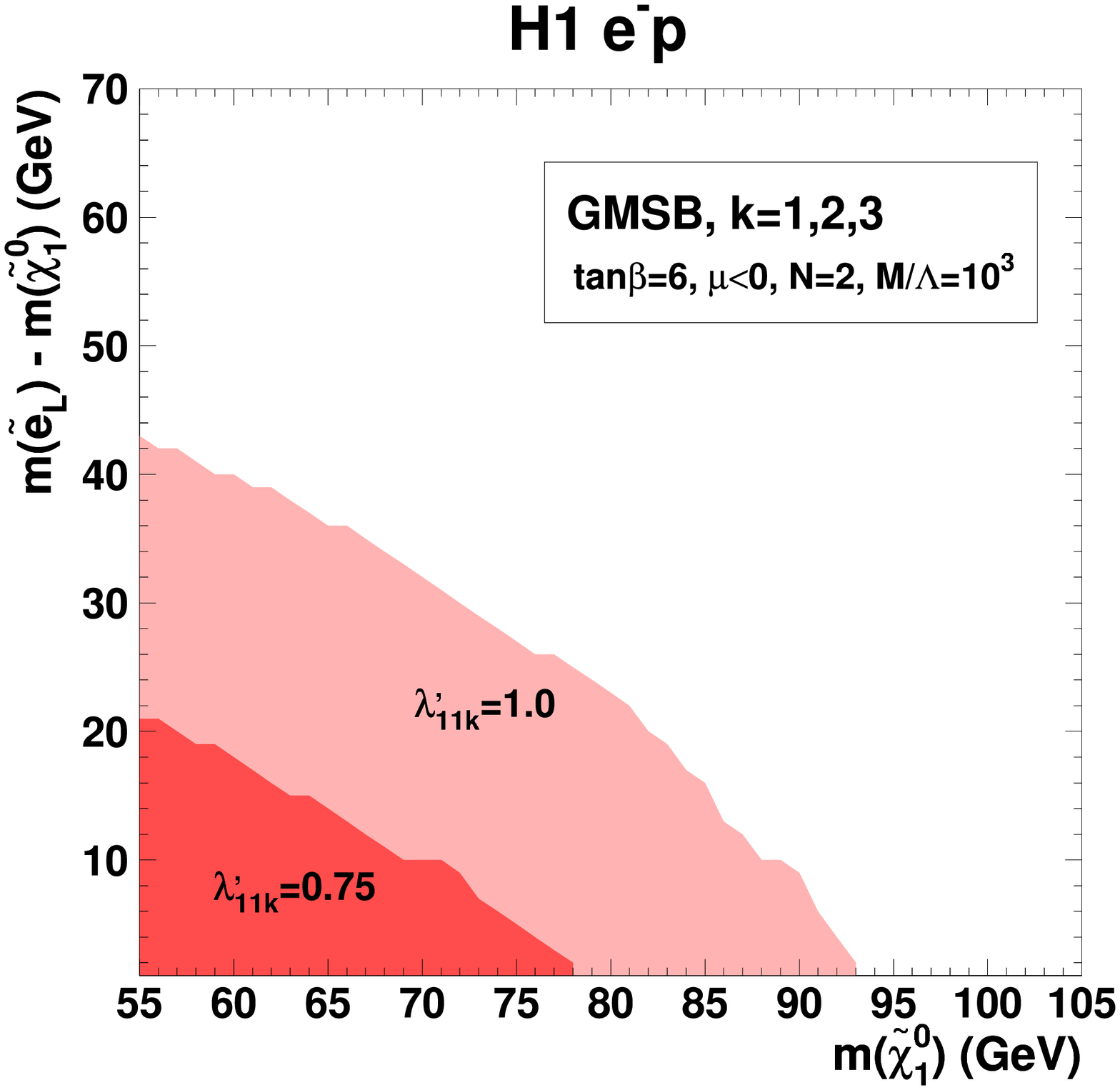}
   \\(c)\hspace{60.6mm}(d)
  \end{center}
  \caption{
   Exlusion regions at $95\%$ CL in the $\Delta m$-$m(\tilde{\chi}_1^0)$ plane.\protect\\
   Examples are shown for $e^+p$-scattering (a,c) and $e^-p$-scattering (b,d) with two different sets of SUSY parameters (a,b and c,d) and varying values of $\lambda'$.
   The signal efficiency is $10\,-\,35\%$.
   These are the first constraints from HERA on SUSY models, which are independent on the squark sector.
  }
  \label{plot_H1_gravitino_Deltam}
 \end{figure}

 In the search performed by the H1 collaboration, the signal has been simulated with varying values of $\tan\beta$, $\Delta m$, $\lambda'$, $N$ and $M/\Lambda$.
 A parameter scan has been performed for $50\,$GeV$\le m(\tilde{\chi}_1^0)\le 140\,$GeV and $m(\tilde{\chi}_1^0)\le m(\tilde{e}_L)\le 200\,$GeV.
 No deviations from the Standard Model have been observed and limits on the scanned parameter space are derived as shown in Fig. \ref{plot_H1_gravitino_Deltam}.

\begin{figure}[htbp]
  \begin{center}
   \includegraphics[width=64.19mm]{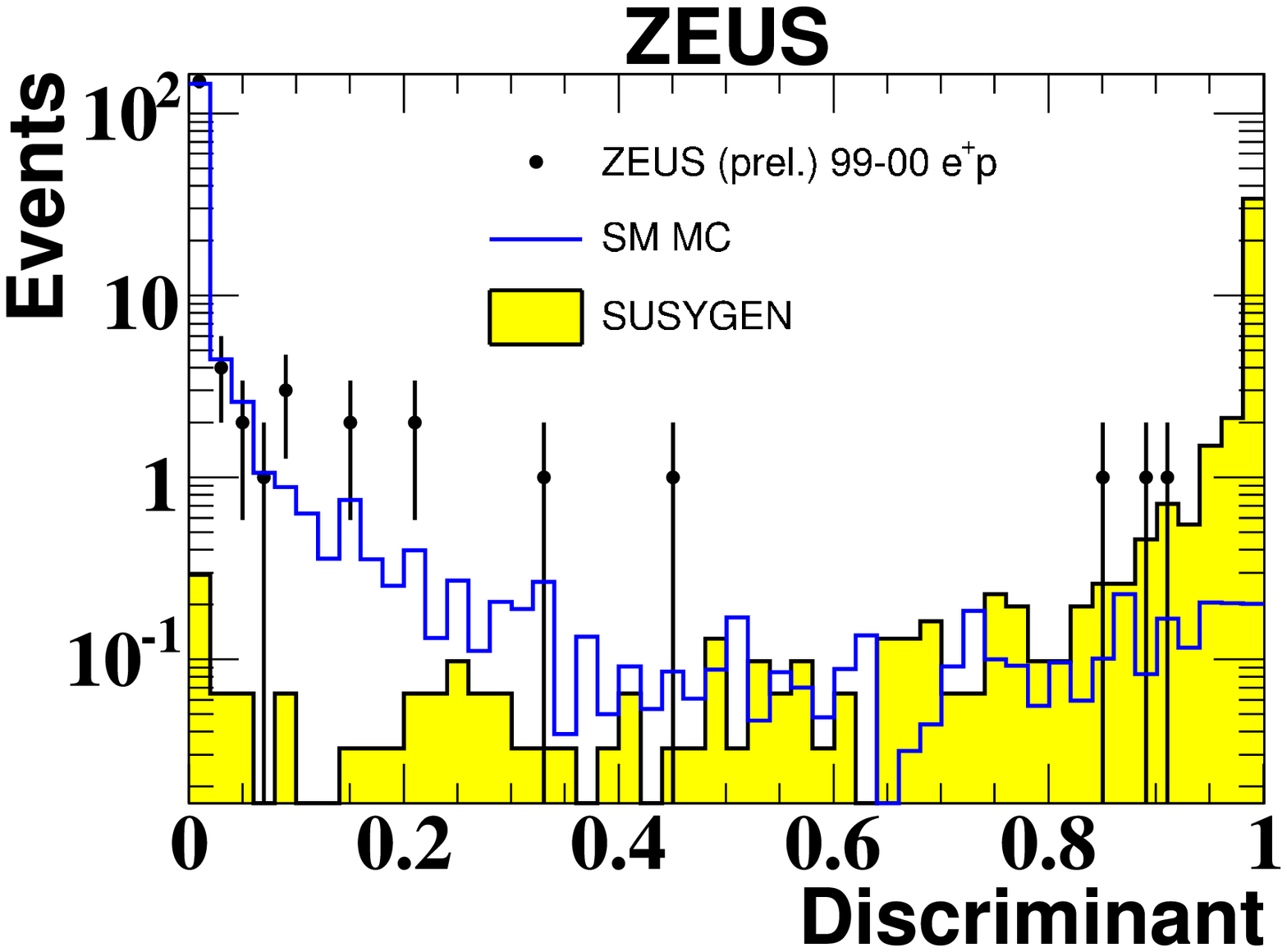}
   \includegraphics[width=65.81mm]{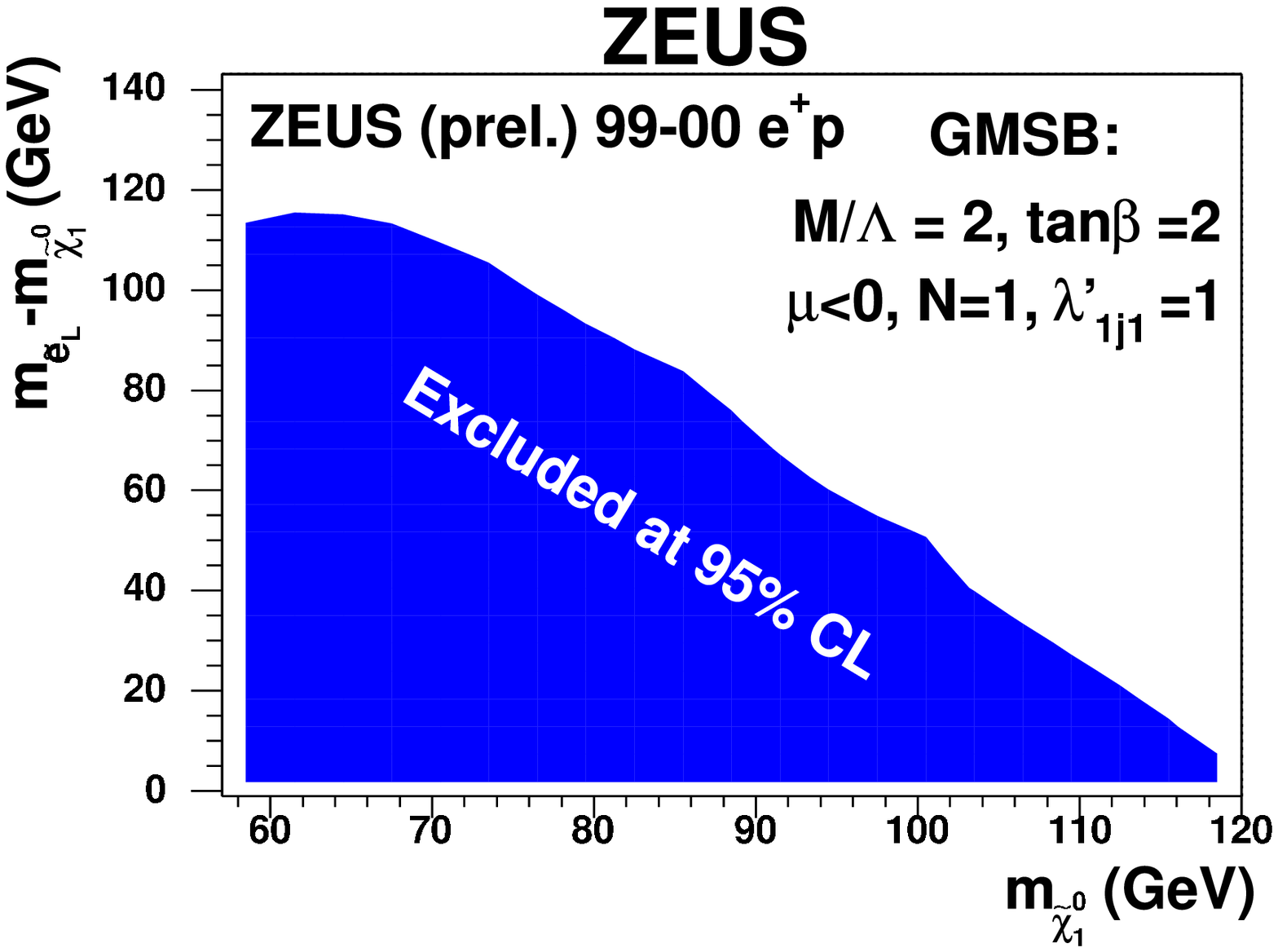}
   \\(a)\hspace{62mm}(b)
  \end{center}
  \caption{
   Discriminant and exlusion region at $95\%$ CL in the $\Delta m$-$m(\tilde{\chi}_1^0)$ plane.\protect\\
   The distribution of the discriminant (a) is shown for the example of $m(\tilde{\chi}_1^0)=100.4\,$GeV and $\Delta m=40\,$GeV.
   No cut on the discriminant is applied.
   The signal efficiency is $59\,-\,71\%$.
   The exclusion limits (b) are derived for one particular set of SUSY parameters.
  }
  \label{plot_ZEUS_gravitino}
 \end{figure}

 In the search performed by the ZEUS collaboration, the signal has been simulated with one set of SUSY parameters in order to optimise the efficiency for this particular set.
 A parameter scan in the $\Delta m$-$m(\tilde{\chi}_1^0)$ plane has been performed with $59\,$GeV$\le m(\tilde{\chi}_1^0)\le 114\,$GeV.
 The found data events have been classified using a multi-variate discriminant method \cite{Carli:2002jp} with six variables.
 As shown in Fig. \ref{plot_ZEUS_gravitino}, no deviations from the Standard Model have been observed in the high-discriminant region and limits on the scanned parameter space are derived.

 \section{Summary}

 Searches for signatures of RPV SUSY have been performed by the H1 and ZEUS collaboration.
 No deviations from the Standard Model have been observed.
 In a search for gaugino production in MSSM, ZEUS extends the existing limits from LEP in the $M_2$-$\mu$ plane.
 In a search for gravitino production in GMSB, H1 sets limits on $m(\tilde{\chi}_1^0)$, $\Delta m$ and $\lambda'$ for various sets of parameters.
 For small mass differences between the neutralino and the selectron, $m(\tilde{\chi}_1^0)<112\,$GeV is ruled out at $95\%$ CL for $\lambda'=1$.
 Similarly, for small masses of the neutralino $m(\tilde{e}_L)<164\,$GeV is excluded.
 Yukawa couplings down to electromagnetic strength ($\lambda'_{1j1}=0.3$) are excluded for both masses close to $55\,$GeV.
 In a complementary search, the mass limits are extended by the ZEUS collaboration for one particular parameter set with $\lambda'=1$.


\begin{thebibliography}{99}
  \bibitem{Nilles:1983ge}
   H.~P.~Nilles,
   Phys.\ Rept.\  {\bf 110} (1984) 1;\protect\\
   H.~E.~Haber and G.~L.~Kane,
   Phys.\ Rept.\  {\bf 117} (1985) 75.
  \bibitem{Fourletova:2005}
   J.~Fourletova,
   {\em Search for RPV Squark Production at HERA},
   in these proceedings,
   PoS(HEP2005)340.
  \bibitem{Carli:2002jp}
   T.~Carli and B.~Koblitz,
   Nucl.\ Instrum.\ Meth.\ A {\bf 501} (2003) 576
   [hep-ex/0211019].
  \bibitem{Aktas:2004cc}
   H1 Collab., A.~Aktas {\it et al.},
   Phys.\ Lett.\ B {\bf 616} (2005) 31
   [hep-ex/0501030].
 \end{thebibliography}
\end{document}